\documentclass[10pt,twocolumn]{article}
\usepackage[textwidth=7in,textheight=8.5in]{geometry}
\usepackage{graphicx}
\usepackage{amsmath} 
\usepackage{amssymb} 
\usepackage{fancyvrb} 
\usepackage{soul}
\usepackage{fancyhdr}
\usepackage{epstopdf}
\usepackage{lineno}
\usepackage{cancel}
\usepackage{subfigure}
\usepackage{hyperref}
\makeatletter
\renewcommand{\maketitle}{\bgroup
\begin{flushleft}
  \begin{Huge}
  \textbf{\@title}\\
  \end{Huge}
  \vspace{1cm}
  \@author
\end{flushleft}\egroup
}
\makeatother
\title{Language Independent Speech Emotion and Non-invasive Early Detection of Neurocognitive Disorder}
\author{%
    \textbf{{\large Susmita Bhaduri}}$^{1}$, \textbf{{\Large Anirban Bhaduri}}$^{2}$, \textbf{{\Large Rajib Sarkar}}$^{3}$\\
    $^{1}$Moonshot Analytics and Design 3/1/1, Dhakruia Station Lane, Kolkata - 700031, West Bengal, India \\
    $^{2}$Moonshot Analytics and Design 3/1/1, Dhakruia Station Lane, Kolkata - 700031, West Bengal, India \\
    $^{3}$Rajarhat Road, R. Gopalpur Kolkata -700136,West Bengal, India\\
    \underline{$^{1}$\textbf{ORCID}:0000-0003-1246-9124, E-mail: susmita.sbhaduri@moonshot.net.in}\\
    \underline{$^{2}$\textbf{ORCID}:0000-0002-7787-3550, E-mail: bhaduri.anirban@moonshot.net.in}\\
    \underline{$^{3}$\textbf{E-mail:} rjbsarkar@gmail.com}\\
}
\begin{document}
\twocolumn[
  \begin{@twocolumnfalse}
    \maketitle
  \end{@twocolumnfalse}
  ]
\noindent

\date{\today}
%
\begin{abstract}
Emotions(like fear,anger,sadness,happiness etc.) are the fundamental features of human behavior and governs his/her mental health. The subtlety of emotional fluctuations can be examined through perturbation in conversations or speech. Analysis of emotional state of a person from acoustical features of speech signal leads to discovery of vital cues determining his/her mental health. Hence, it’s an important field of research in the area of Human Computer Interaction(HCI). In a recent work\cite{Bhaduri20161} we have shown that how the contrast in Hurst-Exponent calculated from the non-stationary and nonlinear aspects of \textit{angry} and \textit{sad} speech(spoken in English language) recordings in the \textit{Toronto-Emotional-Speech-Set(TESS)}\cite{dupuis2010toronto} can be used for early detection and diagnosis of Alzheimer's Disease. In this work we have extended the work and extracted Hurst-exponent for the speech-signals of similar emotions but spoken in German language(\textit{Emo-DB}\cite{burkhardt-2005}). It has been observed that the Hurst-exponent efficiently segregates the contrasting emotions of \textit{anger} and \textit{sadness} in the speech spoken in German language, in similar fashion it has been doing for English speech. Hence it can be concluded that the Hurst-exponent can differentiate among speech spoken out of different emotions in language-independent manner. We propose algorithm for a language-independent application for early non-invasive detection of various severe neurocognitive-disorders like Alzheimer's Disease, MND(motor-neuron-disorder), ASD(autism-spectrum-disorder), depression, suicidal-tendency etc. which is not possible with the state of the art medical science.
\end{abstract}
%

\textbf{PCAS Nos.:} \textit{10, 11.30.Pb, 24.60.Ky, 24.60.Lz}

\section{Introduction}
\label{intro}

Emotions are tangible expression of brain activity finely expressed through speech. Other than speech they are expressed via facial distortions, physical gestures and so on. Speech emotions are expressed both voluntarily or involuntarily. But recognizing the subtle content of emotion in human speech is one  of  the  latest  challenges  in  speech  processing~\cite{el-2011}. Besides human facial expressions, speech has proven as one of the most promising modalities for the automatic recognition of human emotions. Speech emotion recognition process revolves around the speech signal processing to detect and highlight the non-verbal features of speech. The data analysis process based of the parameters related to speech emotion is extremely helpful in various fields like psychology, medical science, lie detection tool, video games and security systems and so on. 

Speech of any language is essentially form of expression of a person's emotion based on the physical and mental situation he or she may be in. Hence, emotion elicited in speech does not depend on the language~\cite{Maganti2014}. Emotion is entwined with the personality, nature, temperament etc. of a person, which again changes from situation to situation.
Neurocognitive disorders involve deterioration of cognitive abilities like memory, problem solving, perception, judgement etc. These disorders result from temporary or permanent damage of the brain, from degenerative processes like  Alzheimer’s or Parkinson's disease, motor neuron disorder dementia and also from affective disorders like depression, suicidal tendency, pathological anxiety and even from bipolar disorder, autism and dyslexia~\cite{Ganguli2011}. 

Among the various types of dementia, noways, most alarming one for elderly people is Alzheimer's Disease with symptoms like irreversible and growing cognitive deterioration, gradual memory loss, weakened judgment, speech impairment among other cognitive deficiencies. Eventually the patient becomes incapable of leading normal professional and social life. The symptoms are normally moderate during the early stage and normally related to age related issues like infrequent memory disorders, impairment in speech, sentence construction etc. Naturally $2-3$ years are spent to take medical step after the beginning of more severe symptoms, that too after expensive neuro-psychological diagnosis. This is a major challenge faced for early detection of any neurocognitive disorder in a non-invasive manner for people in different countries and speaking different languages.

Traditionally different emotions were considered as categorically different and mutually exclusive from each other while devising models for emotion categorization. According to Cornelius \textit{The Big Six Emotions} are: \textit{fear, anger, sadness, happiness, surprise and disgust}~\cite{Cornelius1995}. But later as it was established that emotion is a continuous process, several two-dimensional models have been proposed of which Russell's~\cite{russell1980} and Thayer's~\cite{thayer1990biopsychology} are widely used which are defined by valence (whether the emotion elicits positivity or negativity) and arousal (activity, intensity of the emotion). The obvious criticism of this model is that starkly varying emotions with respect to cognitive, semantic and psychological significance associated with different languages may be close and indistinguishable in this rigidly defined emotional space. 

There have been attempts to analyze various stationary aspects(like the power spectrum~\cite{Hargreaves1965}, energy distribution~\cite{Tolkmitt1982}, prosody, formants, and glottal ratio spectrum~\cite{Sun2011} of speech signals to define non-invasive techniques of early detection of neurocognitive disorders, as speech is the most unsolicited, natural and instinctive means of communication and so it defines one's responsive and cognitive ability. Moreover, apart from Alzheimer's Disease, other types of dementia and motor neuron disorders, other neurocognitive disorders stated above, affect the tonal quality, energy-content impaired expression of emotion in speech irrespective of the languages it's spoken in.
Non-invasive methodologies like Automatic Spontaneous Speech Analysis (ASSA), Emotional Response Analysis (ERA) for detecting neurocognitive disorders, are based on stationary features of speech signal like intensity, short term energy, pitch, spectral centroid etc.

Speech contains a large variety of complex sounds with varying temporal grain, periodic and aperiodic components, noise, frequency and amplitude modulations etc. Most of works have been done to devise speech-emotion analysis system using the conventional stationary acoustic features. They deal with time domain features like Zero Crossing Rate, Short Time Energy and frequency domain features like signal bandwidth, spectral centroid, signal energy, fundamental frequency, Mel-Frequency Cepstral Co-efficients(MFCC). Linear discriminate classifier~\cite{roy-1996}, the k-nearest neighbor (k-NN)~\cite{averill-1994} and Support Vector Machine(SVM)~\cite{you-2006} have mostly been used in this context. Most of these stationary techniques involve Fourier spectral analysis which is based on linear super-positions of trigonometric functions. Secondary harmonic components which is common in natural non-stationary time-series, may generate a distorted wave outline for these natural signals. These distortions are the consequence of nonlinear contributions which are not normally extracted from the non-stationary signals, when analyzed using these stationary techniques. Moreover, none of these attempts are language independent and largely depends on local dialects of various languages. 

All the organs of human body behave nonlinearly due to their inherent complex dynamic nature. The process of speech production and cognition by human beings is complex phenomena~\cite{VanZandt2008}. 
The theory of complexity is rooted in chaos theory~\cite{Poincare1889} and have various parameters whose combined behavior refers to a border between order and randomness, termed as \textit{the edge of chaos}~\cite{Horgan1995}. 
As per chaos theory, a chaotic system is extremely sensitive to initial conditions, does not repeat itself and however is deterministic. The chaos-based complexity theory attempts to decode behavior of dynamic nonlinear systems~\cite{Richard1999,Mikulecky2001,Higgins2003}.

To provide order or definite properties to an structural form inherent in the chaotic system, \textit{fractal geometry} has been evolved~\cite{Peitgen2004}. According to Mandelbrot~\cite{mandelbrot-1968,mandel83}, \textit{fractal} is a geometric scheme which repeats itself at smaller or larger scales to generate \textit{self-similar}, irregular shapes or surfaces that can not be represented by Euclidean geometry. \textit{Fractal} systems can extend to infinitely large values of their coordinates, in all direction from the center towards the outside. The principal feature of \textit{fractal}-s is its \textit{self-similarity}. It is a phenomenon where smaller and bigger fragments of a system looks very alike, but not necessarily exactly the same, to the whole \textit{fractal} system. \textit{Power-law}(as per statistics, a \textit{power-law} is a functional relationship between two quantities, where one quantity varies as a power of another) is applied to represent the self-similarity of the large and small fragments of a \textit{fractal} system. This \textit{power-law} exponent is defined as the \textit{scaling exponent} of the self-similarity or the \textit{fractal dimension} of the system.

Detrended Fluctuation Analysis(DFA) method\cite{Peng1994,Kantelhardt2001} has been implemented for analyzing the long-range correlations in noisy and non-stationary time series and determining their \textit{mono}fractal scaling exponents.
If the time series is long-range correlated, its DFA function shows a \textit{power-law} relationship with its scale parameter. If we denote the DFA function of the time series by $F(s)$ and its scale parameter by $s$, $F(s)$ will vary with a power of $s$ as per the equation $F(s)\propto s^H$, here the exponent $H$ is termed as \textit{Hurst exponent}. If $D_F$ is the \textit{fractal dimension}, it is related with $H$-Hurst exponent as per the equation $D_F=2-H$ \cite{Mandelbrot1985}.
Results obtained by DFA method are proved to be more reliable compared to the methods like Wavelet Analysis, Discrete Wavelet Transform, Wavelet Transform Modulus Maxima, Detrending Moving Average, Band Moving Average, Modified Detrended Fluctuation Analysis etc.~\cite{Oswicimka2006,Serrano2009,Huang2011}. We have applied this method successfully for analysing various kinds of time series formed from natural signals like speech signals\cite{Bhaduri20161} and biological signals like EEG and ECG signals~\cite{Bhaduri2014,Bhaduri2016,P2016,Bhaduri2017}.

The speech production process exhibits \textit{fractal} characteristics. The quasi-static oscillations of the vocal folds and the adaptation process of the vocal tract are both nonlinear processes~\cite{Levelt1999}. Fractal nature of speech has been explored for automatic speech recognition~\cite{Maragos1999a}, speaker recognition~\cite{Gonzalez2012}, speech decomposition~\cite{Langi1997}, speech segmentation, representation and characterization~\cite{Kinsner2008}. Regarding the speech-emotion recognition from nonlinear perspective, classifiers like Artificial Neural Network (ANN) and decision trees are implemented due to their consistent performances in specific cases~\cite{schuller-2007}. However, it has been detected that the same feature vector produces completely different classification performance using varried algorithms~\cite{kim-2013}. Moreover, extraction and selection of features is a crucial parameter in creating language independent speech emotion-detection model and in determining on what emotion a specific speech should be tagged to. Other characteristic features should cover speaking/listening process, linguistics specific to a language and so on. 

Hence, we should define a speech-emotion classification system by analysing speech as complex system using state of the art methods in \textit{fractal} domain, in contrast with the conventional stationary techniques. This way all aspects of speech signal can be understood at the deepest level. In our earlier work~\cite{Bhaduri20162}, we have applied multi-fractal method to the time-displacement profile of speech(non-musical), drone(periodically musical) and Indian art music samples having different musicality(emotions) and showed that the value of the width of the multi-ractal spectrum is substantially different for speech and music signals. 

In another work~\cite{Bhaduri20161}, we have applied the similar approach over speech signal and proposed a quantitative parameter for categorizing various emotions based on the Hurst exponent, by analyzing the non-stationary details of the dynamics of speech signal, generated out of differing emotions. A non-invasive system has been proposed using this parameter for early detection of Alzheimer's Disease. Complex network based \textit{Visibility Graph} method has been applied to analyze the audio signals of speech and music(drone) and a non-invasive model for early detection and monitoring of autism spectrum disorder has been proposed using non-stationary acoustic cues for differentiating speech and music signals in~\cite{Bhaduri2018}. \textit{Modified Visibility Graph analysis} has been done over speech signals spoken out of various emotions and application for detecting suicidal tendency using nonlinear and non-stationary aspects of speech is proposed in~\cite{bhaduriJneuro2016}. However all these applications are designed to work for speech spoken in English language.
In this work we have extended the work and extracted Hurst exponent for the speech signals generated out of similar emotions but spoken in German language. It has been observed that the Hurst Exponent efficiently segregates the contrasting emotions of \textit{anger} and \textit{sadness} in the speech spoken in German language, in a similar fashion it has been doing for the speech signal of English language. Hence it can be concluded that the Hurst exponent can differentiate among speech spoken out of different emotions in a language independent manner. 

Based on the nonlinear and non-stationary parameters of speech(indepenedent of the language it's spoken in) and earlier works~\cite{Bhaduri20162,Bhaduri20161,bhaduriJneuro2016,Bhaduri2018}, we can model non-invasive language independent applications for early detection and constant monitoring of various neurocognitive disorders which would have substantially lesser infrastructural and technological cost. 

This would be a positive step towards defining various applications based on speech emotions in terms nonlinear and non-stationary acoustic parameters of speech signals irrespective of languages they are spoken in. This would make the applications language independent and hence portable.

The rest of the paper is organized as follows. The details of data is in Section~\ref{data}, the method of analysis is elaborated and the inferences from the test results are presented in Section~\ref{analysis}. The inferences are discussed and an example of the application for early detection and monitoring a neurocognitive disorder in non-invasive and language independent manner is elaborated in Section~\ref{con}.

\section{Data}
\label{data}
In this work, we have used two datasets consisting of emotional speech-audio clips, namely {\it Emo-DB}~\cite{burkhardt-2005} for German language and {\it TESS}~\cite{dupuis2010toronto} for English(USA) language.

The recordings of \textit{Emo-DB}~\cite{burkhardt-2005} dataset were done in the anechoic chamber with high quality recording equipment at the Technical University Berlin, department of Technical Acoustics. Ten German emotional utterances, mostly used in everyday communication and could be interpreted in all the applied emotions, were produced in German language by ten actors. The actors(five men and five women) simulated the emotions. The age of all actors were between $21-35$ years. The recorded speech signal-files of {\it Emo-DB} dataset are in .wav format with sampling rate of $16$kHz and of single channel. Duration of the audio files are around $2$ to $5$ seconds. In this experiment, we have taken $127$ recorded speech signal-files spoken out of \textit{anger} emotion and $62$ files created out of \textit{sad} emotion. 

The recordings of Toronto Emotional Speech Set (TESS)~\cite{dupuis2010toronto} dataset created by Kate Dupuis and M. Kathleen Pechora-Fuller at the University of Toronto Psychology Department. We have obtained it from the website:\href{https://tspace.library.utoronto.ca/handle/1807/24487}{TESS-data}.
In \textit{TESS} dataset, there are voice samples of $26$ and $64$ year old actresses speaking a set of $200$ target words in English. The recorded speech signal-files are in .wav format with sampling rate of $24.4$kHz and of single channel. In this experiment, we have taken $200$ recorded speech signal-files spoken out of \textit{anger} emotion and $200$ files created out of \textit{sad} emotion. Moreover, we have considered the speech signal-files created for young actresses because the actors of \textit{Emo-DB} dataset are of the same range of age. 

The amplitude waveforms of the sound files are taken for the experiment. Empirical mode decomposition method~\cite{Huang1998} is applied over the original signal to remove noise. Then each of the .wav file is converted to a corresponding text file.
\section{Method}
\label{meth}
For each of the text files obtained from the .wav file as per the process described in the Section~\ref{data}, Hurst exponenet values are calculated as per the method of Peng et.al.~\cite{Peng1994} and Kantelhardt et.al.~\cite{Kantelhardt2001}. This method is described in Section~\ref{hurst_meth}. For the files created from emotional English speech-signals, two sets of $200$ Hurst exponents are calculated for each of the \textit{angry} and \textit{sad} emotions. For the files created from German recordings $127$ and $62$ Hurst exponents are calculated for \textit{angry} and \textit{sad} emotions respectively. After that with all the Hurst exponent values frequency histogram is created in Section~\ref{analysis} and Hurst exponent values for \textit{anger} and \textit{sad} emotional speech spoken in English and German language are deduced and parameters for language independent non-invasive application for diagnosis of neurocognitive disorders are defined.

\subsection{Hurst Exponent}
\label{hurst_meth}
\begin{figure*}[t]
	\centering
	\includegraphics[width=5in]{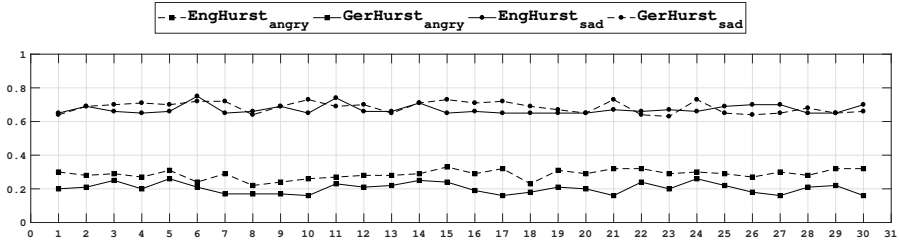}
	\caption{Comparison of the Hurst exponents calculated for $30$ samples for each of the \textit{anger} and \textit{sad} emotions for both the English and German languages.}
	\centering
	\label{indi_hurst}
\end{figure*}

Detrended Fluctuation Analysis(DFA) defined by Peng et.al.~\cite{Peng1994} has been applied successfully over various non-stationary time series for detecting long-range correlations.
\begin{enumerate}
\item Let us denote the input data series as $x(i)$ for $i = 1,2,\ldots,N$, with $N$ number of points. The mean values of this series is calculated as $\bar{x} = \frac{1}{N}\sum_{i=1}^{N} x(i)$. Then accumulated deviation series for $x(i)$ is calculated as per the below equation.
\begin{eqnarray}
X(i) \equiv \sum_{k=1}^{i} [x(k)-\bar{x}], i = 1,2,\ldots,N  \nonumber 
\end{eqnarray}
This subtraction of the mean($\bar{x}$) from the data series, is a standard way of removing noisy data from the input data series. The effect of this subtraction would be eliminated by the detrending in the fourth step.

\item $X(i)$ is divided into $N_s$ non-overlapping segments, where $N_s \equiv int(N/s)$, $s$ is the length of the segment. In our experiment $s$ varies from $16$ as minimum to $1024$ as maximum value in log-scale.

\item For each $s$, we denote a particular segment by $v$($v = 1,2,\ldots,N_s$). For each segment least-square fit is performed to obtain the local trend of the particular segment~\cite{Peng1994}.
Here $x_v(i)$ denotes the least square fitted polynomials for the segment $v$ in $X(i)$. $x_v(i)$ is calculated as per the equations $x_v(i) = \sum_{k=0}^{m} {C_{k}}{(i)^{m-k}}$, where $C_{k}$ is the $k$th coefficients of the fit polynomials with degree $m$. For fitting linear, quadratic, cubic or higher $m$-order polynomials may be used~\cite{Kantelhardt2001,kant2002}. For this experiment $m$ is taken as $1$ for linear fitting.

\item To detrend the data series, we have to subtract the polynomial fit from the data series. There is presence of slow varying trends in natural data series. Hence to quantify the scale invariant structure of the variation around the trends, detrending is required.
Here for each $s$ and segment $v \in 1,2,\ldots,N_s$, detrending is done by subtracting the least-square fit $x_v(i)$ from the part of the data series $X(i)$, for the segment $v$ to determine the variance, denoted by $F^2(s,v)$ calculated as per the following equation.
\begin{eqnarray}
F^2(s,v) \equiv \frac{1}{s}\sum_{i=1}^{s} \{X[(v-1)s+i]-x_v(i)\}^2, \nonumber
\end{eqnarray}
where $s \in 16,32,\ldots,1024$ and $v \in 1,2,\ldots,N_s$.

\item Then the $q$th-order fluctuation function, denoted by $F_q(s)$, is calculated by averaging $F^2(s,v)$ over all the segments($v$) generated for each of the $s \in 16,32,\ldots,1024$ and for a particular $q$, as per the equation below.
\begin{eqnarray}
F_q(s) \equiv \left\{\frac{1}{N_s}\sum_{v=1}^{N_s} [F^2(s,v)]^{\frac{q}{2}}\right\}^{\frac{1}{q}}, \nonumber
\end{eqnarray}
for $q \neq 0$ because in that case $\frac{1}{q}$ would blow up. For $q = 2$, calculation of $F_q(s)$ boils down to standard method of Detrended Fluctuation Analysis(DFA)~\cite{Peng1994}.

\item The above process is repeated for different values of $s \in 16,32,\ldots,1024$ and it can be seen that for a specific $q$, $F_q(s)$ increases with increasing $s$. If the series is long range power correlated, the $F_q(s)$ versus $s$ for a particular $q$, will show power-law behavior as below.
\begin{eqnarray}
F_q(s) \propto s^{h(q)} \nonumber
\end{eqnarray}
If this kind of scaling exists, $\log_{2} [F_q(s)]$ would depend linearly on $\log_{2} s$, where $h(q)$ is the slope which depends on $q$. $h(2)$ is similar to the well-known \textbf{Hurst exponent}~\cite{Kantelhardt2001}. So, in general, $h(q)$ is the generalized Hurst exponent.

\end{enumerate}
If $h(2)=0.5$ then there is no correlation. Further, if $h(2) > 0.5$ then there is persistent long-range cross-correlations, where the large value of one variable is likely to be followed by a large value of another variable in the series, whereas, $h(2) < 0.5$ then there is anti-persistent long-range cross-correlations, where a large value of one variable is most likely to be followed by a small value and vice versa, in the series.
In the Figure~\ref{indi_hurst}, the comparison of the Hurst exponents calculated for $30$ samples for each of the \textit{anger} and \textit{sad} emotions for both the English and German languages are shown. It's evident that the speech signals spoken out of \textit{anger} emotion has anti-persistent long-range correlations, whereas, those of \textit{sad} emotion have persistent long-range correlations, for both the languages.

\subsection{Analysis of Hurst Exponents}
\label{analysis}
Hurst exponents are calculated for
\begin{itemize}
\item Emotional English speech-signals
\begin{enumerate}
\item $200$ Hurst exponent values for speech-signal files recorded for \textit{anger} emotion.
\item $200$ Hurst exponent values for speech-signal files recorded for \textit{sad} emotion.
\end{enumerate}
\item Emotional German speech-signals
\begin{enumerate}
\item $127$ Hurst exponent values for speech-signal files recorded for \textit{anger} emotion.
\item $62$ Hurst exponent values for speech-signal files recorded for \textit{sad} emotion.
\end{enumerate}
\end{itemize}

\begin{figure*}[t]
	\centering
	\includegraphics[width=5in]{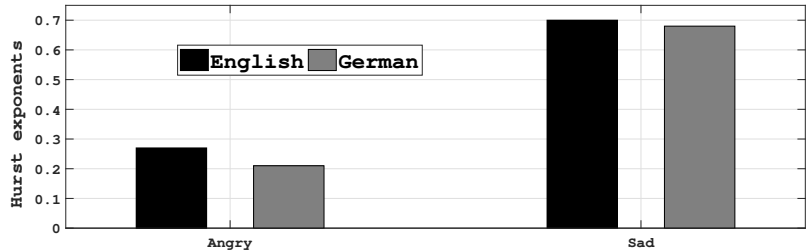}
	\caption{Trend of Hurst exponents for \textit{anger} and \textit{sad} emotion for both the English and German languages.}
	\centering
	\label{hurst}
\end{figure*}

Then for each of the languages, the frequency histogram of occurrence of a specific Hurst exponent for each of the two emotions are formed. The bin width of the histograms is decided on the basis of the range of the Hurst exponents in a particular category of emotion for each of the languages. The peak of a particular histogram formed for the particular set of Hurst exponents calculated for a particular emotion and language, is considered as the Hurst exponent calculated for that emotional speech in that language. Figure~\ref{hurst} shows the comparison of Hurst exponents extracted from the histograms for each emotion and language. The Figure shows that Hurst exponent calculated for \textit{anger} emotion is startlingly less than \textit{sad} emotion, consistently for both the languages.

\section{Conclusion}
\label{con}
The most striking finding of present investigation based on robust nonlinear methods is that the fundamental emotions like \textit{anger and sadness} are clearly distinguishable from the speech signals in a \textbf{language independent manner}, which was not reported in the earlier studies. As the emotions expressed in speech varies from subject to subject a range of values or threshold for the Hurst exponent calculated for each of the emotions of \textit{anger} and \textit{sadness} would be defined. If the Hurst exponent calculated for an emotional speech signal spoken in any language falls within any of the two ranges(for \textit{anger} or \textit{sadness}) then we can detect and tag the emotion of the speech as any of the two. 
This range should be defined by analyzing the fractal properties of time-displacement profile using the method in Section~\ref{hurst} and Section~\ref{analysis}, for large number of speech samples of different emotions and different language. Using this method, non-invasive applications for detecting various severe neorocognitive disorders like Alzheimer’s disease, motor neuron disorder, autism spectrum disorder, disorder of consciousness, suicidal tendency etc., can be devised. As an example, we have broadly outlined a framework for one such application for early detection of neorocognitive disorders in the Section~\ref{ex}.

\subsection{\label{ex}Proposed Exemplary Application}
\label{ex}
As already mentioned in the Section~\ref{intro}, neurocognitive disorders involve cognitive impairment restricting proper emotional expression in speech, memory problem, issues involving problem solving, perception, judgment etc. and these disorders result from temporary or permanent damage to the brain, degenerative processes like Alzheimer’s or Parkinson's disease, dementia, motor neuron disorders and also from affective disorders like depression, pathological anxiety and even from bipolar disorder, autism and dyslexia~\cite{Ganguli2011}. Recently we have proposed a quantitative parameter for categorizing various emotions based on the Hurst exponent, by analyzing the non-stationary details of the dynamics of speech signal spoken in English language, generated out of differing emotions. A non-invasive system has been framed using this parameter for early detection of Alzheimer's Disease~\cite{Bhaduri20161}. Further we have applied complex network based \textit{Visibility Graph} and \textit{Modified Visibility Graph} methods to analyze non-stationary aspects of speech(spoken in English language) and proposed non-invasive application for early detection of autism spectrum disorder and suicidal tendency~\cite{Bhaduri2018,bhaduriJneuro2016}. 
The initial version of the proposed algorithm for speech emotion detection is in the website:\href{http://www.dgfoundation.in/speech-emotion/}{Speech-Emotion}.
The present experiment confirms that the Hurst exponent is language independent for emotional(\textit{anger} and \textit{sadness}) speech signals.

In this work, we propose a quantitative framework to capture the change in intricate dynamics of speech spoken in any language by normal subject and a subject suffering from neurocognitive disorders. According to the steps elaborated below, three types of control elements are calculated. 
\begin{itemize}
\item One is for normal subjects
\item Second for subjects who have already been diagnosed with some kind of neurocognitive disorder
\item Third one is for any subject to be diagnosed for similar disorder. 
\end{itemize}
Then depending upon the proximity of the third one to the first or second one, proneness or onset of disorder can be decided. 

\begin{enumerate}
\item First emotional speech signals generated by large number of normal subjects speaking varying languages, would be collected. Here we have considered the two most fundamental emotions of \textit{anger} and \textit{sadness}.Then, after doing the scaling analysis of amplitude-profile of the speech signals using the method described in Section~\ref{meth} the range of Hurst exponents for normal subjects, say denoted by $H_{norm}\pm\delta_{norm}$, can be base-lined. This would be first control element for this application for monitoring and early detection of any neorocognitive disorder.

\item Similarly, the audio clip of the emotional speech signals generated by the subjects already been diagnosed with any of the  neorocognitive disorders, would be recorded. People suffering from neorocognitive disorders elicit emotion in their speech differently than normal, irrespective of the language they are speaking in. Hence, the speech spoken out of fundamental emotions by such subjects, would definitely display different scaling behavior in all acoustic aspects, than those of normal subjects. Therefore, using the same method of Step 1, the second control element, say denoted by $H_{dis}\pm\delta_{dis}$, for diseased subjects suffering from neorocognitive disorders, can be base-lined.

\item Different ranges of the first two control elements with varying $\delta_{norm}$ and $\delta_{dis}$, would be defined, to reflect the proneness or the severity of neorocognitive disorders. One sample set of ranges is given below.
\begin{enumerate}
\item \textit{Severity 1:}First range for deciding that whether the subject to be diagnosed is at all prone to neorocognitive disorders or not.
\item \textit{Severity 2:}Second range for deciding the onset of neorocognitive disorders.
\item \textit{Severity 3:}Third for prognosis of neorocognitive disorders.
\end{enumerate}

\item Simple and lightweight android application would be framed where the two types of control elements(calculated for normal and diseased subjects) containing ranges of language independent Hurst exponents with varying $\delta$ values calculated from emotional speech signals for fundamental emotions, would be stored locally.

When the subject would be speaking over the phone, his/her speech signal would be recorded in .wav format and the Hurst exponent would be calculated from its amplitude-profile. This would be the third control element as stated before. This would be matched against the first two control elements, and according to the severity ranges of $\delta_{norm}$ and $\delta_{dis}$ defined in Step 3 to decide the proneness to any neorocognitive disorder, is decided for the subject. The application may be scheduled to be run as per the default frequency preset in the application or it might run as a background application and the third control element would be calculated each time the subject would make a phone call. Third control element would be checked against the \textit{Severity 1}-range, which would signify the proneness of the subject to the neorocognitive disorder. 

If this range of values is obtained for the third element for more than certain present frequency, and the third control element is getting nearer to \textit{Severity 2}-range an alarm would be generated and the onset of the disorder can be confirmed and accordingly further prognosis may be started. This would be a real-time, non-invasive, language independent and routine check-up framework for self-assessment as well as monitoring of any neurocognitive disorder.

The \textit{Severity 3}-range would be decided based on the prognosis parameters. Once the onset of the disease is confirmed the subject would be under necessary treatment and then the third control element would be matched against the first two control elements according to this severity range. This way continuous monitoring of the patient would be done during prognosis period until the proximity with \textit{Severity 1}-range would be met, which would in turn set a guideline for the treatment of the patient.

\item This would be definitely an cost effective application and an efficient one because the control elements are defined on the basis of the fundamental and most prominent emotions of \textit{anger} and \textit{sadness}, so the subjects speaking any language would certainly elicit these emotions most of the time. And, as we have already discussed, the less prone a subject would be to any neurocognitive disorder the more prominently the fundamental emotions would be expressed in his/her speech and vice-verse. So, the control elements defined based on these emotions would be the most effective ones and the scaling disparity between normal and person prone to these disorders, would enable us in the most robust and efficient way, to detect the disorder in non-invasive and language independent manner.

\item For the first two control elements, feedback system would be implemented as more and more number of speech samples of normal and diseased subjects speaking various languages would be analyzed for emotion detection, and the control elements would be revised accordingly and eventually the performance of the android application would be refined.
 
\end{enumerate}

\end{document}